\documentclass[fleqn,11pt]{wlscirep}
\usepackage[utf8]{inputenc}
\usepackage[T1]{fontenc}

\usepackage{setspace}
\usepackage{lineno}
\usepackage{siunitx}

\title{Hierarchical resonators for localized charging in room-scale magnetoquasistatic wireless power systems}

\author[1,$\dag$]{Takuya Sasatani}
\author[2]{Alanson P. Sample}
\author[1]{Yoshihiro Kawahara}
\affil[1]{Department of Electrical Engineering and Information Systems, The University of Tokyo, Tokyo, Japan.}
\affil[2]{Electrical Engineering and Computer Science Department, University of Michigan, Ann Arbor, USA.}
\affil[$\dag$]{Corresponding author: \href{mailto:sasatani@g.ecc.u-tokyo.ac.jp}{sasatani@g.ecc.u-tokyo.ac.jp}}
\begin{abstract}
\end{abstract}
\begin{document}

\flushbottom
\maketitle

\thispagestyle{empty}
\vspace{-20pt}
\section*{Abstract}
Magnetoquasistatic wireless power transfer can deliver substantial power to mobile devices over near-field links. Room-scale implementations, such as quasistatic cavity resonators, extend this capability over large enclosed volumes, but their efficiency drops sharply for centimeter-scale or misoriented receivers because the magnetic field is spatially broad and weakly coupled to small coils. Here, we introduce hierarchical resonators that act as selectively activated relays within a room-scale quasistatic cavity resonator, coupling to the ambient magnetic field and re-emitting a stronger local field near a target receiver. This architecture enables localized power delivery to miniature devices without requiring global reshaping of the room-scale cavity mode. Experimentally, the hierarchical link improves power transfer efficiency by more than two orders of magnitude relative to direct room-scale transfer and delivers up to 500 mW of DC power to a 15 mm receiver. We further demonstrate selective multi-relay operation and field reorientation for furniture-embedded charging scenarios. These results establish a scalable route to reconfigurable wireless power delivery for miniature and batteryless devices in room-scale environments.

\section*{Introduction}
The vision of the Internet of Things (IoT) has been significantly advanced by the development of ubiquitous sensing systems enabled by wireless communication, but is now fundamentally constrained by how distributed electronic devices are powered~\cite{Xu2014, Weiser1999, huang_wireless_2019, portilla_wirelessly_2023}. As we transition toward spaces populated by hundreds of miniature sensors and actuators, traditional power solutions—wired connections and batteries—pose significant bottlenecks due to frequent manual recharging or battery replacement, as well as their environmental impact. Wireless power transfer offers a pathway to untether these devices, yet a critical gap remains: safely and efficiently delivering power to small-scale receivers across large volumes~\cite{Hui2014}.

Traditionally, wide-range wireless power solutions rely on radiative electromagnetic fields, such as microwaves~\cite{Shinohara2011, Brown1984, Strassner2013,li2024intelligent,wang_high-performance_2023}. While effective over several meters, these approaches require receiver sizes comparable to the operating wavelength to maintain power efficiency~\cite{garnica_wireless_2013} and are fundamentally limited by the significant interaction with biological tissue~\cite{protection_guidelines_1998, ICNIRP2020}. These constraints often result in deliverable power levels that are insufficient for many IoT nodes. The use of standing waves generated by cavity resonators has been explored but faces similar limitations due to electric field exposure~\cite{chabalko_resonant_2014,tamura_cavity-resonance-enabled_2026}. Magnetoquasistatic approaches provide a safer, high-power alternative~\cite{Kurs2009, Sample2011, Assawaworrarit2020,song_wireless_2021,chen_modelling_2024}; because magnetic fields at MHz or lower frequencies interact minimally with biological tissue, they allow for significantly higher power levels while remaining within safety limits~\cite{Christ2013, hirata2026electromagnetic}. However, conventional systems are typically limited to short-range transfer, making charging pads or surfaces the primary use case~\cite{Hui2014,sumiya_alvus_2019}.

To extend the range of magnetoquasistatic wireless power transfer, room-scale systems based on quasistatic cavity resonators (QSCRs) have emerged as a promising approach~\cite{Chabalko2017, sasatani_geometry-based_2022, sasatani_room-wide_2018}. These structures use conductive surfaces and lumped capacitors to generate homogeneous magnetic field patterns that permeate large enclosed volumes. By confining electric fields to discrete capacitive elements, they can deliver tens of watts throughout a room while remaining within established safety guidelines~\cite{Chabalko2017, Sasatani2021, hirata2026electromagnetic}. However, a fundamental scaling barrier remains: in highly asymmetric systems, power transfer efficiency drops sharply as the receiver size decreases. Prior work extended QSCRs to multimode structures that support multiple magnetic field patterns and improve coverage, providing a degree of reconfigurability, but generating localized fields on demand remains challenging~\cite{Sasatani2021, Sasatani2017}.

This scaling challenge is fundamentally rooted in the physics of highly asymmetric coupled-resonator systems. The maximum efficiency $\eta_{\rm max}$ achievable between a transmitter and receiver is determined by the dimensionless coupling parameter $k\sqrt{Q_1Q_2}$, hereafter referred to as the $kQ$-product, where $k$ is the coupling coefficient and $Q_1$ and $Q_2$ are the quality factors of the transmitter and receiver, respectively~\cite{Zargham2012}. In the weak-coupling regime characteristic of miniature receivers operating within room-scale transmitters, this relation simplifies to $\eta_{\rm max} \approx \frac{1}{4} k^2 Q_1 Q_2$. Because the coupling coefficient $k$ is proportional to the magnetic flux linkage, and that flux linkage scales with receiver area $A \propto d_{\rm RX}^2$, where $d_{\rm RX}$ is the receiver's characteristic dimension, the resulting efficiency $\eta$ scales as $d_{\rm RX}^4$ or more steeply. This scaling law imposes a severe penalty on miniaturization, because the homogeneous fields of room-scale systems lack the spatial localization needed to maintain sufficient flux linkage for centimeter-scale nodes.

Here, we demonstrate a hierarchical resonator architecture that uses intermediate relay resonators as local coupling interfaces between a room-scale transmitter and miniature receivers (Figures~\ref{figure:overview}a-c). Rather than globally reshaping the cavity mode, an activated relay couples to the ambient room-scale magnetic field and re-emits a stronger local field near the receiver. This localized field intensification increases flux linkage to centimeter-scale receivers, while relay orientation can also be used to reorient the local magnetic-field axis for misaligned devices. Integrating these relays into furniture or architectural elements is particularly effective because the built environment often defines where devices are placed and how they are oriented. In this work, we experimentally characterize this architecture and show that the hierarchical link increases power delivery efficiency for pebble-sized receivers by more than two orders of magnitude over direct transfer, delivering up to $\qty{500}{mW}$. The relay state can be controlled using bistable electronic switches, enabling on-demand local power delivery with negligible static power overhead (Figure~\ref{figure:overview}d). This system-level architectural approach offers a pathway for selective yet ubiquitous power delivery, supporting the deployment of power-hungry, small IoT devices in complex smart environments.

\section*{Results}
\subsection*{Scaling principle and hierarchical resonator architecture}
To characterize the factors that limit asymmetric wireless power systems—from room-scale transmitters to miniature receivers—we first examine the coupling physics governing these disparate scales. As illustrated in Figure~\ref{figure:overview}e, the maximum efficiency $\eta_{\rm max}$ achievable between a pair of resonators with quality factors $Q_1$ and $Q_2$ and coupling coefficient $k_{1,2}$ is determined by the dimensionless parameter $k_{1,2}\sqrt{Q_1Q_2}$ (the $kQ$-product)~\cite{Zargham2012,Haus1991}:

\begin{equation}
\eta_{\rm max} = \dfrac{k_{1,2}^2 Q_1 Q_2}{\left(1+\sqrt{1+k_{1,2}^2 Q_1 Q_2}\right)^2}
\label{equation:max_eff}
\end{equation}

The coupling coefficient $k_{1,2}$ is physically related to the magnetic flux linkage $\beta$ through the receiver and the total magnetic energy stored in the cavity volume $\alpha$~\cite{chabalko_resonant_2014}:

\begin{equation}
k_{1,2} = \dfrac{\beta}{\sqrt{2L_2\alpha}}
\label{eq:hierarchicalWPT:kappa}
\end{equation}
\begin{equation}
\beta = \iint_{A_2} \left(\mu_0\vec{H}_1\cdot\vec{n}\right)\,\mathrm{d}A_2
\label{eq:hierarchicalWPT:beta}
\end{equation}
\begin{equation}
\alpha = \iiint_{V_1} \left(\mu_0\dfrac{|\vec{H}_1|^2}{2}\right)\,\mathrm{d}V_1
\label{eq:hierarchicalWPT:alpha}
\end{equation}

When the receiver is much smaller than the transmitter volume, the system operates in the weak-coupling regime ($kQ \ll 1$). Under this condition, the maximum efficiency simplifies to
\begin{equation}
\eta_{\rm max} \approx \frac{1}{4} k^2 Q_1 Q_2
\end{equation}
Because the magnetic field $\vec{H}_1$ generated by a room-scale transmitter is approximately uniform across the aperture $A_2$ of a miniature receiver, the flux linkage scales directly with area, $\beta \propto A_2 \propto d_{\rm RX}^2$, where $d_{\rm RX}$ denotes the receiver's characteristic dimension. Consequently, as shown in Figure~\ref{figure:overview}e, the efficiency scales as $d_{\rm RX}^4$ or more steeply. This scaling law imposes a severe penalty on miniaturization: a system capable of delivering high efficiency to mobile-sized ($\sim\qty{100}{mm}$) receivers can fall below $0.1\%$ efficiency when scaled down to tiny ($\sim\qty{10}{mm}$) or misaligned nodes. This sharp decline fundamentally limits direct power transfer to miniature devices. Our hierarchical architecture addresses this constraint by introducing relay resonators that locally concentrate the magnetic field near the receiver.

\subsection*{Localized field intensification with hierarchical resonators}
The magnetic flux linkage of miniature or misaligned receivers can be enhanced by locally increasing the magnetic field amplitude or changing its vector orientation near the receiver. Our approach uses relay resonators that couple to the ambient room-scale magnetic field and re-emit a secondary magnetic field in the vicinity of the resonators. The resulting field is a superposition of the original transmitter field and the localized relay field, producing a local high-coupling region rather than reshaping the global cavity mode.

When embedded into elements of the built environment, these relay modules provide local power-delivery interfaces at selected device locations. The primary energy source remains the room-scale transmitter, so the relay action itself is passive once activated.

\begin{figure}[t]
\centering
\includegraphics[width=0.96\linewidth]{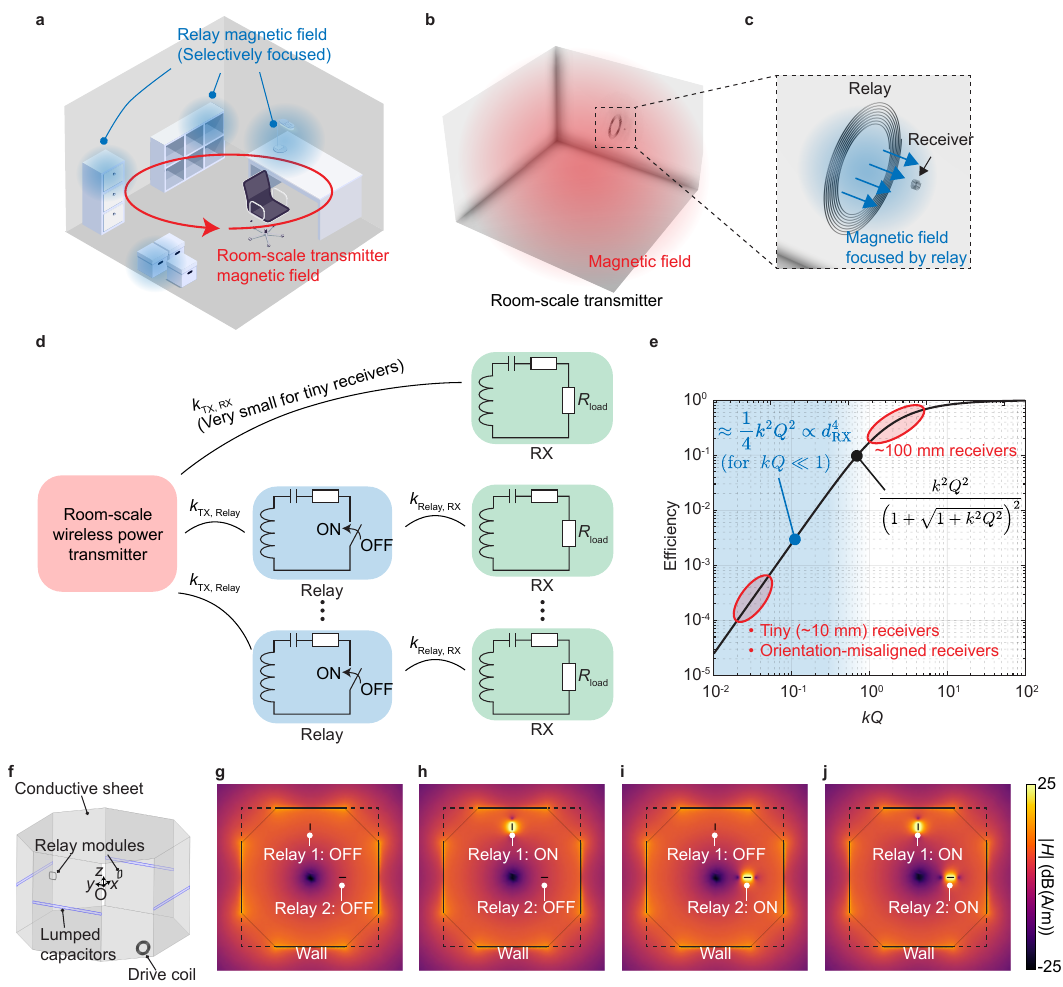}
\caption{\textbf{Concept of hierarchical resonators for wireless power transfer.}
\textbf{a} Representative deployment scenario: the room-scale transmitter generates a wide-area magnetic field, while interior objects embed relay modules that form hierarchical links and selectively focus the field.
\textbf{b, c} Field-focusing effect, in which intermediate relay resonators concentrate and redirect magnetic flux from the transmitter to miniature receivers.
\textbf{d} Hierarchical coupling architecture. Because direct transmitter--receiver coupling ($k_{\rm TX,RX}$) is very weak, hierarchical links enhance power transfer efficiency.
\textbf{e} Power transfer efficiency versus $kQ$-product, with representative operating points for standard ($\sim100~\mathrm{mm}$) and tiny ($\sim10~\mathrm{mm}$) or misaligned receivers. In the weak-coupling regime, efficiency scales approximately as $d_{\rm RX}^4$, where $d_{\rm RX}$ is the receiver's characteristic dimension.
\textbf{f} Three-dimensional model of the cavity resonator; conductive sheets with integrated lumped capacitors generate the magnetic field while confining electric fields. The origin $O$ denotes the center of the structure; the drive coil is located at $(x,y,z)=(-0.9,-1.35,-0.8)\,\mathrm{m}$ in the simulations and is placed approximately at this position in the experiments, since the precise drive-coil placement has little effect on the relay-induced field (Fig.~S1).
\textbf{g} Top-down baseline field distribution with both relays inactive, showing a homogeneous pattern.
\textbf{h, i} Reconfigured field distributions with Relay 1 and Relay 2 individually activated, respectively.
\textbf{j} Field distribution with both relays activated; the localized amplification near the active relays illustrates selective power direction within large spaces.
The furniture illustrations in panel \textbf{a} are adapted from a stock illustration \copyright~iStock.com/aklionka (image ID 1005460100), used under an iStock standard content license; these elements are not covered by the article's Creative Commons license.
}
\label{figure:overview}
\end{figure}

Furthermore, the relay effect can be toggled on demand via electronic switches integrated into the resonator loop. By utilizing bistable components—such as latched mechanical relays~\cite{sumiya_alvus_2019}—the module requires energy only during state transitions and consumes zero static power to maintain its active state. This enables efficient, on-demand power distribution with negligible power overhead. Figure~\ref{figure:overview}f-j illustrates the simulated magnetic field patterns generated by this hierarchical approach. Using a 3-D model of a room-scale transmitter built on quasistatic cavity resonance technology~\cite{Sasatani2021}, simulations show that the field distribution is approximately homogeneous when relays are inactive (Figures~\ref{figure:overview}f-g). This baseline distribution exhibits the nearly uniform magnetic field characteristic of direct power transfer. In contrast, activating a relay significantly intensifies the local field in its proximity, while the global field remains nearly stable (Figures~\ref{figure:overview}h,i). When multiple relays are activated simultaneously, the localized enhancement is distributed among the active relay locations rather than concentrated at a single relay (Figure~\ref{figure:overview}j). This selective reconfiguration enables efficient power transfer to previously unreachable receivers using direct-transfer methods.

The drive coil used to excite the cavity mode was placed near the cavity wall, as indicated in Figure~\ref{figure:overview}f. The precise drive-coil orientation is not a dominant factor in the relay-induced reconfiguration pattern, provided that the drive coil maintains sufficient coupling to the desired QSCR mode. Fig.~S1 shows that the relay-assisted efficiency remains nearly constant for drive-coil rotations up to about $60^\circ$, and that varying the drive-coil orientation mainly changes the localized near field around the feed, while the broader room-scale field and relay-induced localized enhancement remain similar. This behavior reflects the superposition of the local drive-coil field and the cavity-mediated QSCR mode.

\subsection*{Evaluation of power transfer efficiency}

\begin{figure}[t]
\centering
\includegraphics[width=\linewidth]{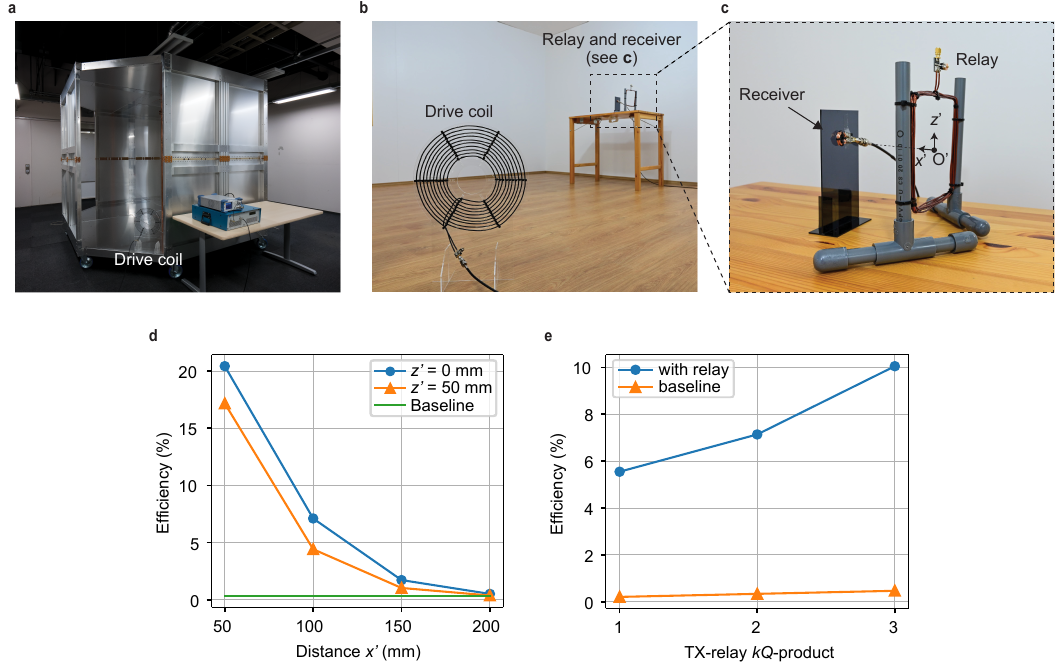}
\caption{\textbf{Experimental setup for efficiency measurements.}
\textbf{a} Photograph of the room-scale quasistatic cavity resonator used in this study, adapted from our prior work~\cite{Sasatani2021}.
\textbf{b} Interior view of the furnished cavity, showing the drive coil and a representative relay--receiver measurement arrangement.
\textbf{c} Detailed view of the relay module and miniature receiver, together with the local coordinate system used to define receiver position relative to the relay.
\textbf{d} Power transfer efficiency as a function of the distance between the relay and receiver for two vertical offsets ($z^\prime=0$ and $z^\prime=\qty{50}{mm}$), compared with the baseline direct-transfer case.
\textbf{e} Power transfer efficiency as a function of the transmitter--relay $kQ$-product, showing a substantial improvement of the hierarchical relay link over baseline direct transfer.}
\label{figure:evaluation}
\end{figure}

We evaluated the hierarchical wireless power system using a $\qty{3}{m} \times \qty{3}{m} \times \qty{2}{m}$ room-scale transmitter excited by a drive coil, consistent with the configuration established in prior work (Figures~\ref{figure:evaluation}a,b)~\cite{Sasatani2021}. The system was operated in the pole-independent (PI) mode because of its broad volumetric coverage, although the proposed approach is not fundamentally restricted to this mode. To express relay location in a setup-independent manner, we use the $kQ$-product of the transmitter--relay link as a measure of effective coupling, while receiver position is specified using a local coordinate system ($x^\prime, y^\prime, z^\prime$) centered on the relay module (Figure~\ref{figure:evaluation}c). Although the QSCR field is approximately homogeneous over the room-scale volume, the transmitter--relay coupling is not identical for all relay placements. The coupling coefficient is determined by the magnetic flux linked through the finite relay aperture and therefore depends on the relay position, orientation, and local placement conditions. In the experiments, these physical variations were summarized by the measured transmitter--relay $kQ$-product, $k_{\mathrm{TX,relay}}\sqrt{Q_{\mathrm{TX}}Q_{\mathrm{relay}}}$. Thus, the $kQ$-product should be interpreted as a scalar descriptor of the effective relay coupling to the cavity mode, rather than as a uniquely defined spatial coordinate.

All resonant components were tuned to a center frequency of $\qty{1.34}{MHz}$. The relay module consisted of a $\qty{150}{mm}$ square, 6-turn coil, whereas the receiver used a $\qty{15}{mm}$-diameter, 10-turn coil. The quality factors of the drive coil, room-scale transmitter, relay resonator, and receiver coil were measured as $Q_{\rm drive}=396$, $Q_{\rm TX}=540$, $Q_{\rm relay}=202$, and $Q_{\rm RX}=53$, respectively~\cite{kajfez_q-factor_1984}. For these initial benchmarks, the relay was activated by a direct short connection to evaluate the maximum achievable relay effect.

S-parameters between the drive coil and the $\qty{15}{mm}$ receiver were measured using a vector network analyzer (VNA). The maximum power transfer efficiency was then calculated assuming the use of appropriate impedance tuning~\cite{Zargham2012,Sample2013,Sasatani2021,Chabalko2017}. This S-parameter-based value represents the AC-to-AC efficiency of the passive resonant wireless link under optimized impedance tuning and is distinct from the end-to-end DC-to-DC efficiency reported below, which additionally includes the transmitter and receiver power electronics used in the prototype.

Figure~\ref{figure:evaluation}d shows the effect of receiver position relative to the relay, with the transmitter--relay coupling fixed at a $kQ$-product of 2. Without a relay, the baseline direct-transfer efficiency remains low at approximately $0.3\%$ across the measured range. When the relay is activated, the efficiency increases to more than $20\%$ at a relative distance of $\qty{50}{mm}$ ($z^\prime=0$). This enhancement persists under vertical misalignment, and even at $z^\prime=\qty{50}{mm}$, the efficiency remains well above the baseline value.

\begin{figure}[t]
\centering
\includegraphics[width=\linewidth]{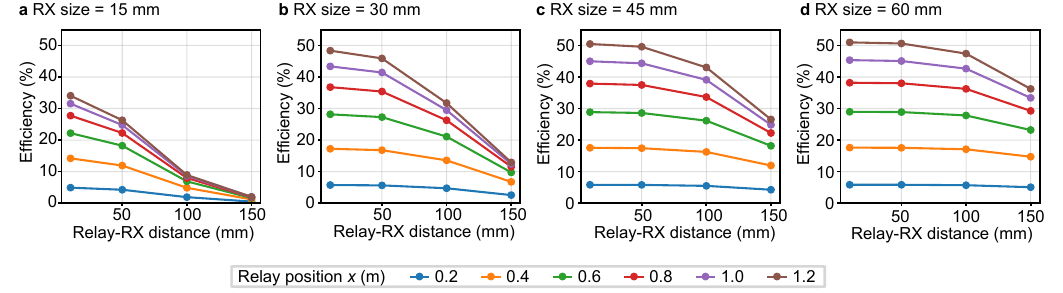}
\caption{\textbf{Simulated power transfer efficiency of the hierarchical standard relay link.}
Simulated power transfer efficiency of the hierarchical relay link as a function of relay--receiver distance, shown for four receiver sizes: (a)~\qty{15}{mm}, (b)~\qty{30}{mm}, (c)~\qty{45}{mm}, and (d)~\qty{60}{mm}. In each panel, color denotes the relay position from the cavity center, swept from \qty{0.2}{m} to \qty{1.2}{m}, which sets the transmitter--relay coupling strength. Relay positions of \qty{0.4}{m}, \qty{0.8}{m}, and \qty{1.2}{m} correspond to nominal transmitter--relay $kQ$-products of 1, 2, and 3, respectively.}
\label{figure:hierarchical-sweeps}
\end{figure}

To examine the effect of relay placement, Figure~\ref{figure:evaluation}e plots efficiency as a function of the transmitter--relay $kQ$-product, with the receiver fixed at $x^\prime=\qty{100}{mm}$ and $z^\prime=0$. Although the baseline efficiency increases slightly with stronger transmitter--relay coupling because of the stronger ambient field, the relay-assisted link retains a clear performance advantage. The efficiency increases from approximately $5.5\%$ at $kQ=1$ to nearly $10\%$ at $kQ=3$. These results show that the hierarchical architecture improves transfer efficiency across a range of spatial configurations and coupling conditions. Complementing these measurements, simulations of the standard relay link across a range of receiver sizes and relay positions reproduce these trends and show how the achievable efficiency scales with receiver size (Fig.~\ref{figure:hierarchical-sweeps}).

In addition to the S-parameter (AC-to-AC) analysis, we further simulated the passive-link power flow under the same representative relay-assisted condition. For this characterization, the relay was positioned such that $k_{\rm TX,relay}\sqrt{Q_{\rm TX}Q_{\rm relay}} = 2$, while the receiver was fixed at $(x^\prime,y^\prime,z^\prime) = (100, 0, 0)~\mathrm{mm}$ relative to the relay. Fig.~S2 shows the simulated power-flow breakdown expressed as a percentage (\%) of the accepted RF power. This accepted RF power (100\%) is either delivered to the receiver load (7.97\%) or dissipated in the drive coil (2.72\%), room-scale QSCR (22.19\%), relay resonator (60.95\%), and receiver coil (6.17\%). This analysis clarifies the loss distribution within the passive hierarchical link, while excluding implementation-specific amplifier and rectification losses. The accepted RF power is consumed predominantly in the relay resonator, which carries the large circulating current required to intensify the local field.

\subsection*{Multichannel operation and system throughput}
\begin{figure}[t]
\centering
\includegraphics[width=\linewidth]{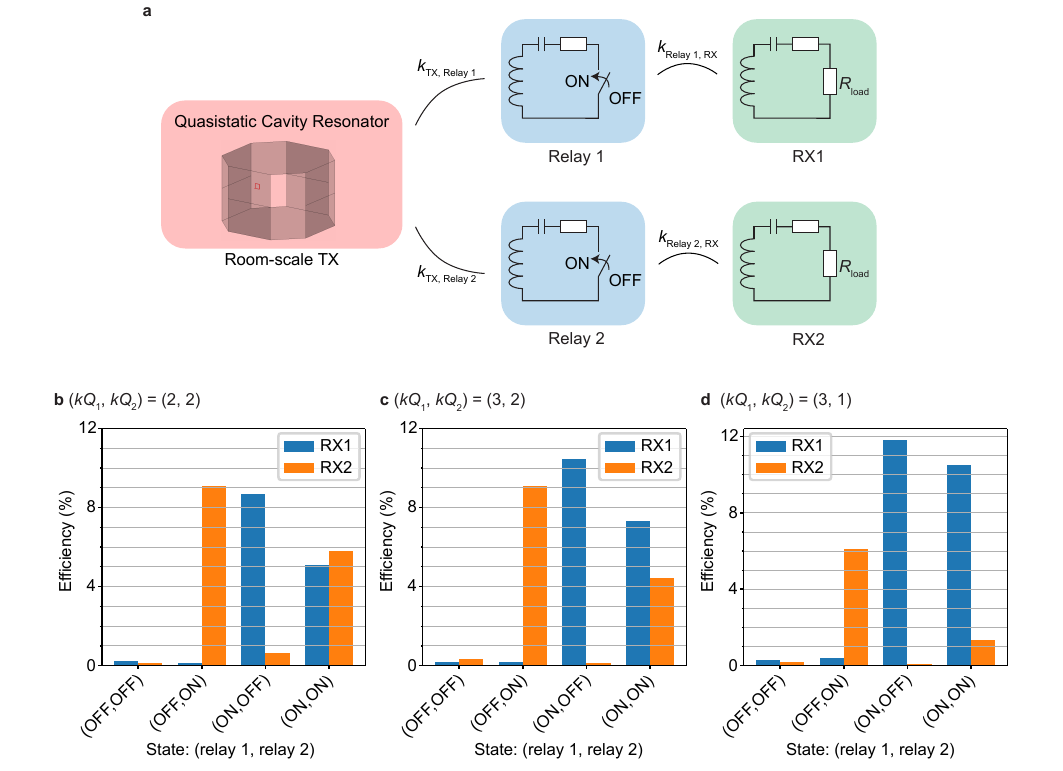}
\caption{\textbf{Multi-relay reconfigurability.}
\textbf{a} Schematic of a room-scale transmitter supporting two hierarchical links, from Relay 1 to RX1 and from Relay 2 to RX2.
\textbf{b--d} Measured power transfer efficiency of RX1 and RX2 for the four relay states (OFF, OFF), (OFF, ON), (ON, OFF), and (ON, ON) under different transmitter--relay $kQ$ configurations. The results show how simultaneous relay activation redistributes power between the two links depending on their relative coupling strengths.}
\label{figure:multiple-relay}
\end{figure}

We further investigated system behavior in the presence of multiple relay--receiver pairs, as illustrated in Figure~\ref{figure:multiple-relay}a. Two relays were positioned at locations corresponding to $kQ$ values of 1, 2, and 3 to emulate both balanced and unbalanced coupling scenarios. For each configuration, efficiency was evaluated by switching each relay between ON (short) and OFF (open) states. The results show that activating multiple relays simultaneously introduces competition for the ambient room-scale field (Figure~\ref{figure:multiple-relay}b--d). When both relays are active (ON, ON), the efficiency of each individual link (RX1 and RX2) decreases relative to the corresponding single-relay states (ON, OFF and OFF, ON). Even so, the efficiency remains substantially above the baseline as long as the corresponding relay is active.

This degradation depends on the relative coupling strengths of the transmitter--relay links. In a balanced scenario ($kQ_1=2$, $kQ_2=2$), both receivers exhibit a moderate and nearly symmetric drop in efficiency. By contrast, in an imbalanced scenario ($kQ_1=3$, $kQ_2=1$), the weaker link (RX2) experiences a larger reduction when Relay 1 is activated, because the more strongly coupled relay draws a disproportionate share of the transmitter energy. To visualize the corresponding field redistribution, Figure~\ref{figure:overview}j shows the simulated magnetic-field distribution when both relays are simultaneously active. Compared with the single-relay cases, the localized field enhancement is split between the active relay locations, consistent with the measured reduction in individual link efficiency in the multi-relay experiments.

To further examine scalability beyond the two-relay configuration, we analyzed a five-relay/five-receiver system using a simplified coupled-circuit model under all relay activation patterns (Fig.~S3). The analysis shows that simultaneous relay activation progressively loads the room-scale resonator and redistributes power among active links, leading to gradual per-node power reduction rather than a sharp universal capacity threshold. Thus, the system does not have a single geometry-independent capacity limit; rather, the number of relays that can be activated simultaneously depends on the transmitter--relay coupling strengths, relay--receiver couplings, activation pattern, and power required by each node. This behavior suggests that dense deployments should use selective or time-multiplexed relay activation according to receiver demand rather than activating all relays concurrently.

\subsection*{DC-to-DC performance}
\begin{figure}[t]
\centering
\includegraphics[width=\linewidth]{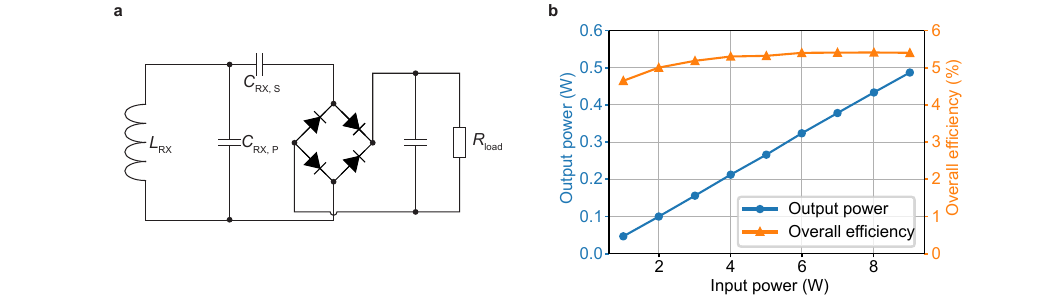}
\caption{\textbf{Rectified power performance and overall efficiency.}
\textbf{a} Circuit schematic of the miniature receiver, including the resonant coil ($L_{\rm RX}$), tuning capacitors, and a full-wave bridge rectifier.
\textbf{b} Measured output power and overall DC-to-DC efficiency as functions of input power. The system delivers up to $\qty{500}{mW}$ of rectified output power while maintaining an overall efficiency of approximately $5\%$.}
\label{figure:dc-to-dc}
\end{figure}

Finally, we characterized the end-to-end DC-to-DC efficiency and deliverable output power. The receiver was integrated with a full-bridge rectifier and tuning capacitors (Figure~\ref{figure:dc-to-dc}a) to deliver power to a $100~\Omega$ DC load. We define the end-to-end efficiency as the ratio of the power delivered to the DC load to the power drawn from the DC source driving the RF power amplifier. For this characterization, the relay was positioned such that $k_{\rm TX,relay}\sqrt{Q_{\rm TX}Q_{\rm relay}} = 2$, while the receiver was fixed at $(x^\prime,y^\prime,z^\prime) = (100, 0, 0)~\mathrm{mm}$ relative to the relay. The impedance-tuning circuit was designed to maximize transfer efficiency under these conditions, resulting in the network topology shown in Figure~\ref{figure:dc-to-dc}a. We report this DC-to-DC value as an end-to-end prototype demonstration, following common practice in resonant wireless power transfer, where the intrinsic wireless link is characterized separately using AC-to-AC efficiency, and DC-to-DC efficiency is used to indicate deliverable rectified power under a specific transmitter and receiver implementation~\cite{9159915,Assawaworrarit2020,Sasatani2021}. We therefore do not interpret the DC-to-DC value as a component-level loss breakdown of the wireless link.

As shown in Figure~\ref{figure:dc-to-dc}b, the overall DC-to-DC efficiency remains approximately constant at $\sim 5\%$ over the measured input-power range. This curve was measured using a fixed impedance-tuning state and therefore includes the power-dependent behavior of the RF power stage, matching network, nonlinear full-bridge rectifier, and load. The tuning capacitor bank was not dynamically adjusted during the input-power sweep. This value is lower than the corresponding AC-to-AC passive-link efficiency, which was approximately $8\%$ under the same spatial conditions, owing to conversion losses in the transmitter and receiver electronics. The system delivered up to $500~\mathrm{mW}$ of DC power to the load with a $9~\mathrm{W}$ input, demonstrating the feasibility of powering miniature IoT devices from a room-scale source through the hierarchical relay architecture.
\subsection*{Demonstration of localized field reconfiguration}
\begin{figure}[t]
\centering
\includegraphics[width=\linewidth]{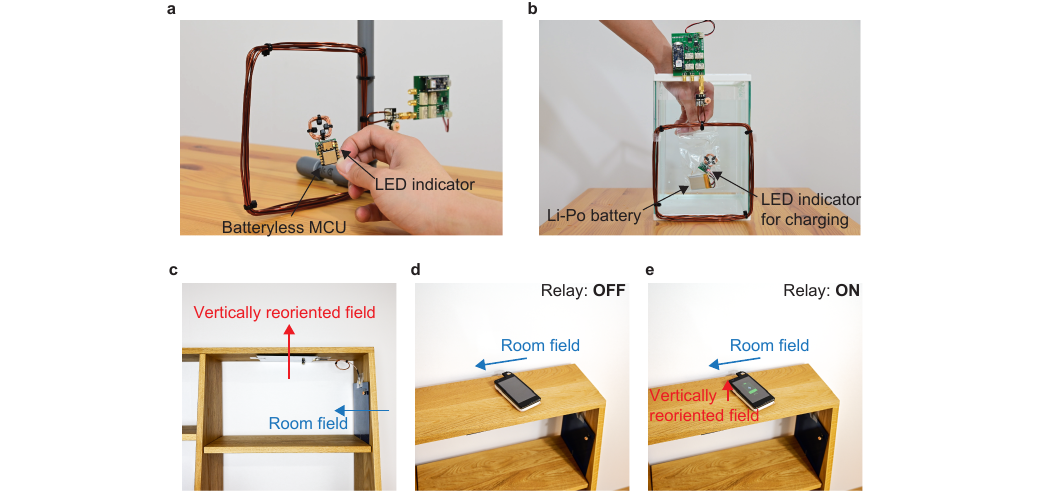}
\caption{\textbf{Demonstrations and usage scenarios of hierarchical resonators.}
\textbf{a} Localized power delivery to a batteryless microcontroller unit (MCU) with an LED indicator.
\textbf{b} Charging a Li-Po battery in water, illustrating the weak interaction of magnetic fields with dielectric environments.
\textbf{c} A field-bending relay module embedded in a wooden shelf, designed to capture the horizontal room field and re-emit it as a vertically reoriented field.
\textbf{d, e} Smartphone charging demonstration showing the transition from no charging with the relay OFF (\textbf{d}) to active charging with the relay ON (\textbf{e}) through field reconfiguration.
Panels \textbf{c}--\textbf{e} are adapted from photographs in the Supplementary Information of our prior work~\cite{Sasatani2021}.}
\label{figure:demonstrations}
\end{figure}

The utility of the hierarchical link was illustrated through several example deployment scenarios (Figure~\ref{figure:demonstrations}). We demonstrated power delivery to batteryless microcontrollers and battery-charging circuits submerged in water, highlighting the weak interaction of magnetic fields with dielectric environments. This supports power delivery to off-the-shelf components without extensive low-power redesign. These relay modules incorporate microcontroller-based control circuits that harvest energy directly from the room's magnetic field to power their own logic, while bistable mechanical relays are used to switch between operational states without static power consumption.

Furthermore, we demonstrated a field-bending function based on the same hierarchical system model, in which a relay captures a horizontal cavity field and re-emits it along a vertical axis (Figures~\ref{figure:demonstrations}c--e). This mechanism addresses orientation-dependent coupling by enabling power delivery to misaligned devices that would otherwise intercept little magnetic flux. We illustrated this concept by embedding the relay into a wooden shelf, a practical placement strategy because devices stored on shelves often have constrained orientations. These results suggest that hierarchical resonators can serve as dynamic interfaces for localized wireless power delivery in built environments.

To quantify the field-bending link beyond the qualitative charging demonstration, we simulated its efficiency across a range of receiver sizes at representative relay positions from the single-relay characterization (\qty{0.4}{m}, \qty{0.8}{m}, and \qty{1.2}{m} from the cavity center, corresponding to $kQ=1,2,3$). Here two wired, orthogonally oriented coil sections form a single relay: one couples to the horizontal cavity field while the other re-emits it toward the receiver along the vertical axis. This split between capturing and re-emitting reduces the effective coupling, so the field-bending link is less efficient than the standard single-relay link at comparable locations. Because direct transfer to an orthogonally misaligned receiver is expected to be near zero (negligible flux linkage), we report the absolute relay-assisted efficiency rather than an improvement ratio; as shown in Fig.~S4, the link still delivers useful efficiency across the tested receiver sizes.

\subsection*{Electromagnetic safety of localized relay fields}
Because the relay resonator locally intensifies the magnetic field, we evaluated electromagnetic safety under the operating condition corresponding to a received power of \qty{500}{mW}. Fig.~S5 shows the simulated external magnetic-field distribution near an active relay, together with dosimetric quantities evaluated using a homogeneous human-equivalent cylindrical phantom~\cite{hirata2013confirmation,ITISDatabase,gabriel1996dielectric}. The magnetic field is strongest immediately adjacent to the relay and decreases rapidly with distance. In this near-field region, the external-field reference level can be locally exceeded; however, this reference level is a conservative screening criterion rather than the underlying safety limit~\cite{ICNIRP2020}.

For the localized RF exposure considered here, the relevant ICNIRP basic restrictions are based on electromagnetic energy absorption in tissue~\cite{ICNIRP2020}. We therefore evaluated peak 10 g SAR and whole-body average SAR. Under the \qty{500}{mW} received-power condition, both SAR metrics remained well below the ICNIRP general-public basic restrictions for the evaluated human--relay distances. These results indicate that the localized field enhancement produced by the relay remains within the relevant exposure limits under the demonstrated operating condition. Because SAR scales approximately linearly with the delivered power in this linear magnetoquasistatic regime, these margins also bound higher-power operation: the maximum compliant received power can be obtained by linear scaling from the demonstrated \qty{500}{mW} condition.

\section*{Discussion}

The hierarchical resonator architecture introduced here addresses the fundamental "coupling gap" inherent in room-scale magnetoquasistatic environments. While traditional room-scale systems provide high power to mobile-sized receivers, they are constrained by a $d_{\rm RX}^4$ efficiency scaling that severely penalizes smaller nodes. Our results demonstrate that semi-passive relay modules can effectively bridge this gap, enhancing power delivery efficiency by over two orders of magnitude and delivering up to $\qty{500}{mW}$ to $\qty{15}{mm}$ receivers.

A direct way to improve coupling in a room-scale magnetoquasistatic link is to increase the receiver coil size. When the target device can accommodate a large resonant receiver, such as a $\qty{150}{mm}$-scale coil, receiver-side enlargement is a valid and potentially preferable solution and may provide higher direct-transfer efficiency than a cascaded relay link under the same placement and orientation conditions. The hierarchical architecture proposed here is intended for the complementary regime in which the receiver size, placement, or orientation is constrained by the device or object being powered. The key distinction is that the relay separates the large magnetic capture aperture from the mobile or embedded receiver. A relay resonator can be integrated into stationary infrastructure such as furniture, shelves, desks, or walls, where dimensions of hundreds of millimeters impose little burden on the powered device itself. The receiver can therefore remain miniature, while the relay provides a local high-coupling interface to the room-scale transmitter.

This separation also provides capabilities that are not obtained by simply enlarging the receiver coil attached to the device. The relay bridges the multiscale coupling gap by coupling more strongly to the room-scale transmitter through its larger aperture and then coupling to the miniature receiver over a short local distance. Because the relay is spatially and electrically decoupled from the receiver, local power delivery can also be selectively activated without modifying the receiver. In addition, the relay can reorient the magnetic-field axis, enabling power delivery to receivers whose placement or orientation would otherwise result in weak flux linkage. Thus, the relay-based architecture should be viewed not as a replacement for large receiver coils in all cases, but as an infrastructure-based approach for devices that must remain small, mobile, embedded, or orientation-constrained.

The relay acts as an infrastructure-side magnetic aperture, so its size should be chosen according to the available space in the surrounding furniture or architectural element, rather than being dictated by the form factor of the powered device. Larger relay coils generally provide stronger coupling to the room-scale cavity field because they intercept more magnetic flux, while larger receiver coils intercept more of the relay-generated local field. Consistent with this scaling, Fig.~S6 shows that the hierarchical-link efficiency increases with both relay size and receiver size at each relay position, and further with stronger transmitter--relay coupling. In practice, the relay should be placed and oriented to maximize flux linkage with the room-scale cavity mode while remaining close to the intended receiver location. The transmitter--relay coupling can be summarized by the measured $kQ$-product, $k_{\mathrm{TX,relay}}\sqrt{Q_{\mathrm{TX}}Q_{\mathrm{relay}}}$, which captures the combined effects of relay position, orientation, and local placement conditions. In this work, relay placements corresponding to $kQ=1$--3 produced substantial efficiency gains for a $\qty{15}{mm}$ receiver, with stronger transmitter--relay coupling generally increasing the relay-assisted efficiency. The receiver should also be positioned within the local near-field region of the relay and oriented to couple to the relay-generated magnetic-field component. For orientation-constrained devices, the field-bending relay provides an alternative configuration that trades some efficiency for magnetic-field-axis reorientation. Overall, the hierarchical architecture is most beneficial in this weak direct-coupling regime, where the receiver is too small, poorly oriented, or too weakly linked to the room-scale field for efficient direct transfer.

In mixed deployments, a field-bending relay can also affect other receivers in two ways: through the superposition of its re-emitted field at the receiver, and through cavity loading. For a horizontally oriented receiver, the vertically re-emitted relay field is orthogonal to the receiver coil axis and links negligible flux, so it neither reinforces nor cancels the flux linked from the background cavity field. The imperfectly vertical components of the relay near field do superpose onto the horizontal cavity field, constructively or destructively, but this is significant only near the edges of the relay coil. Whether a neighboring receiver is affected at all is further set by its separation from the relay: the re-emitted field of an active relay is localized, returning to the background cavity level within roughly \qty{0.2}{m}, comparable to the relay dimension (Fig.~S5a), so a receiver separated by more than about one relay dimension is only weakly affected whatever its orientation; this separation can serve as a practical spacing guideline when embedding relays in furniture or architectural elements, and closer placements may instead call for time-multiplexed activation. Independent of separation and orientation, an active relay also loads the room-scale resonator and redistributes energy among the active links; this loading, rather than the local field superposition, is the dominant influence on other links, and it is managed through the same relay-state control and activation scheduling used for dense multi-relay deployments.

Future research will focus on transitioning from manual relay control to autonomous control logic based on real-time receiver demand and spatial occupancy. More deeply integrating these resonators into the built environment could enable more seamless wireless power delivery. In addition, understanding the collective behavior and potential mutual interference of large numbers of automatically controlled relays will be essential for scaling this architecture to support dense deployments in complex, multi-user smart environments. The multi-relay analysis in Fig.~S3 suggests that system capacity should be treated as a power-allocation and scheduling problem rather than a fixed relay-count limit: the infrastructure may support many deployed relays, while only a subset should be activated simultaneously depending on receiver power requirements and local coupling conditions.

The switching element used to activate each relay is another important implementation parameter. In this prototype, bistable mechanical relays were used because they provide low on-state resistance and consume power only during state transitions, preserving the semi-passive nature of the relay module. At $\qty{1.34}{MHz}$, however, any physical switch introduces parasitic resistance, inductance, and capacitance that can detune the relay resonator or reduce ($Q_{\mathrm{relay}}$). Future implementations could use solid-state RF switches, such as PIN diodes or FET-based switches, to enable faster reconfiguration, smaller form factors, and improved mechanical reliability. These alternatives introduce their own trade-offs, including finite on-state resistance, off-state capacitance, leakage in the inactive state, and possible static bias power. Optimizing the switch technology will therefore be important for balancing switching speed, standby power, reliability, and hierarchical-link efficiency.

\section*{Methods}

\subsection*{Efficiency characterization}
The system's power transfer efficiency was characterized using two-port S-parameter measurements obtained via a vector network analyzer (VNA). The maximum achievable efficiency ($\eta_{\rm max}$) was calculated from the measured S-parameters~\cite{Zargham2012}. This calculation assumes an appropriately designed tuning capacitor bank, providing a benchmark for the theoretical limit of the hierarchical link. The resulting AC-to-AC efficiency is used as the link-level metric for the coupled-resonator link, following common practice in resonant wireless power transfer, where intrinsic wireless-link efficiency is separated from implementation-specific DC-to-RF and RF-to-DC conversion losses~\cite{Zargham2012,Sample2011,Sasatani2021}. A supplementary power-flow analysis further decomposes this AC-to-AC link efficiency into simulated losses in the drive coil, room-scale QSCR, relay resonator, receiver coil, and receiver load. All resonant structures—including the room-scale transmitter, relay modules, and receiver coils—were tuned to a center frequency of $\qty{1.34}{MHz}$ using high-Q capacitors to minimize dissipative losses.

To generalize performance across different spatial configurations within the cavity, the $kQ$-product of the transmitter--relay link was employed as a compact metric for coupling strength. Although the QSCR produces an approximately uniform room-scale magnetic field, the coupling to a finite relay coil depends on the magnetic flux linked through the relay aperture and is therefore affected by relay position, orientation, and local placement conditions. To obtain different transmitter--relay coupling conditions, the relay was physically placed at different locations and orientations inside the cavity. For each placement, the transmitter--relay coupling coefficient was extracted from the measured S-parameters and converted to the $kQ$-product, $k_{\mathrm{TX,relay}}\sqrt{Q_{\mathrm{TX}}Q_{\mathrm{relay}}}$. This procedure allowed position, orientation, and local field variations to be represented by a single effective coupling metric.

The same S-parameter-based maximum-achievable-efficiency calculation was used for both the relay-assisted link and the direct-transfer baseline. Thus, the baseline direct-transfer case was evaluated under the same optimal-matching assumption and was not penalized by a fixed or poorly matched impedance network.

\subsection*{End-to-end measurements}
DC-to-DC efficiency was evaluated by integrating the receiver with a full-bridge rectifier. Efficiency was defined as the ratio of the DC power delivered to a $100~\Omega$ resistive load to the total DC power input to the transmitter drive circuit. Throughput and thermal stability were assessed at input power levels up to $\qty{9}{W}$.

For the batteryless node demonstration, an Atmel ATtiny-series microcontroller equipped with an LED indicator was used to verify functional power delivery within the localized intensification zones. The battery-charging demonstration used a linear charging circuit configured for a constant-current output of $\qty{50}{mA}$. The field-bending capability was demonstrated using a smartphone equipped with a custom-designed receiver coil and a power-management circuit to accommodate variable rectified voltages.

On the transmitter side, a DC power source (Keysight E36312A) supplied power to a high-frequency zero-voltage-switching (ZVS) Class-D power amplifier development board (EPC9065). The amplifier output was coupled to the room-scale transmitter via a drive coil. In evaluating system performance, the auxiliary power consumption of the clock source and logic supplies was excluded, as these represent fixed overhead rather than transfer-related losses. Accordingly, the DC input to the power stage was treated as the primary system input. The output power delivered to the load ($R_{\rm load}$, Figure~\ref{figure:dc-to-dc}a) was recorded using a high-precision digital multimeter.

The impedance-tuning network used for the DC-to-DC measurement was tuned statically before the input-power sweep and left unadjusted during the sweep. The target tuning condition was calculated from the expected AC input impedance of the full-bridge rectifier connected to the $\qty{100}{\ohm}$ DC load. The same tuning state was then used for all input-power levels up to $\qty{9}{W}$. Therefore, the measured efficiency curve in Figure~\ref{figure:dc-to-dc}b includes the effect of power-dependent changes in the rectifier input impedance as well as the transmitter power-stage behavior.

\subsection*{Electromagnetic field simulation}
The RF module of COMSOL Multiphysics was used for the visualization of magnetic field distributions presented in Figures~\ref{figure:overview}g-j. The magnetic field intensity was computed using the Electromagnetic Waves, Frequency Domain (emw) physics interface. The simulation utilized a frequency domain study to determine the field patterns resulting from a normalized net input power of $\qty{1}{W}$, defined as the power actually accepted by the system ($P_{\text{inc}}(1 - |S_{11}|^2)$). This normalization accounts for the operation of high-efficiency switching inverters, where the reflected power does not directly translate into dissipative loss as it might in traditional linear amplifiers used in communication studies. These models account for the localized field intensification and reconfiguration provided by the relay resonators within the room-scale cavity environment.

The drive coil was placed near the cavity wall, consistent with the experimental setup, and oriented to maintain sufficient coupling to the desired QSCR mode. For the drive-coil sensitivity analysis in Fig.~S1, the drive-coil orientation was varied while keeping the cavity and relay geometries fixed.

The simulated efficiency sweeps (Figs.~\ref{figure:hierarchical-sweeps}, S4, and S6) used the same COMSOL RF model. For each configuration, the two-port S-parameters of the drive-coil--receiver link were exported, and the maximum achievable efficiency was computed following the same procedure used for the measured data. In these simulations, the transmitter--relay coupling was varied by placing the relay at different distances from the cavity center, with relay distances of \qty{0.4}{m}, \qty{0.8}{m}, and \qty{1.2}{m} corresponding to nominal transmitter--relay $kQ$-products of 1, 2, and 3, respectively. The quality factor of each component was calibrated to its measured value by inserting a lumped series resistor in the coil models (drive coil, relay, and receiver) and by adjusting the wall conductivity of the room-scale QSCR. Throughout, receiver size denotes the diameter of the circular receiver coil and relay size denotes the edge length of the square relay coil.

The passive-link power flow (Fig.~S2) was extracted from the same electromagnetic model, normalized to the accepted RF power, for the geometry used in the DC-to-DC experiment ($k_{\rm TX,relay}\sqrt{Q_{\rm TX}Q_{\rm relay}}=2$; receiver at $(x^\prime,y^\prime,z^\prime)=(100,0,0)~\mathrm{mm}$ relative to the relay). Ohmic and eddy-current losses were integrated over the conductive elements of the drive coil, room-scale QSCR, relay resonator, and receiver coil, and the load-delivered power was obtained from the load dissipation. This provides an estimate of the AC-to-AC power flow within the passive hierarchical link, complementary to the measured DC-to-DC efficiency, which additionally includes implementation-specific power-electronics losses.

\subsection*{Electromagnetic safety evaluation}

Electromagnetic safety was evaluated for the relay-assisted operating condition corresponding to \qty{500}{mW} received power. For the localized RF exposure considered here, a primary safety endpoint is tissue heating caused by electromagnetic energy absorption. We therefore evaluated the specific absorption rate (SAR), which quantifies absorbed RF power per unit tissue mass, and compared the resulting peak 10 g SAR and whole-body average SAR with the ICNIRP general-public basic restrictions. We also calculated the external magnetic-field distribution around the active relay and compared it with the ICNIRP reference level. The reference level serves as a conservative screening criterion for external fields; if it is locally exceeded, compliance is assessed using the underlying dosimetric basic restrictions, such as SAR.

SAR was evaluated using a simplified homogeneous human-equivalent phantom. Following prior MHz-band wireless-power-transfer dosimetry work, the human body was approximated as a cylindrical phantom \qty{0.23}{m} in diameter and \qty{1.7}{m} in height rather than an anatomically detailed voxel model~\cite{hirata2013confirmation}. The phantom was assigned homogeneous, isotropic dielectric properties equal to two-thirds those of muscle, consistent with prior simplified WPT exposure models~\cite{hirata2013confirmation}. Taking muscle properties at \qty{1.34}{MHz} from the IT'IS/Gabriel tissue-property database~\cite{ITISDatabase,gabriel1996dielectric} and applying the two-thirds factor gives $\epsilon_r=877$, $\sigma=\qty{0.349}{S/m}$, and $\mu_r=1$, with mass density $\rho=\qty{1090}{kg/m^3}$.

The electromagnetic field was solved in COMSOL Multiphysics, and the local SAR distribution was obtained as $\mathrm{SAR} = \sigma|E_{\mathrm{rms}}|^2/\rho$, where $\sigma$ and $\rho$ are the phantom conductivity and mass density. SAR values were scaled to a common received power of \qty{500}{mW}. For the homogeneous phantom, 10 g mass averaging reduces to a moving cubic-volume average with side length $L = (0.01/\rho)^{1/3}$, and the peak 10 g SAR was taken as the maximum of this averaged distribution. The whole-body average SAR was computed by averaging the local SAR over the phantom volume.

\subsection*{Multi-receiver coupled-circuit analysis}

Multi-receiver scalability was analyzed using a frequency-domain impedance-matrix model of the coupled resonators at $f_0=\qty{1.34}{MHz}$. The model comprised the room-scale QSCR transmitter and five relay–receiver pairs, with each resonator represented as a series-tuned RLC branch. The QSCR was driven directly by an ideal voltage source, so efficiencies are reported relative to the power coupled into the QSCR. The diagonal terms of the impedance matrix were set by the self resistances, including receiver loads, while the off-diagonal terms were given by the mutual reactances $j\omega_0 k_{ij}\sqrt{L_iL_j}$. Coil inductances and couplings were obtained from electromagnetic simulation, and the quality factor of each coil was calibrated to its measured value. The QSCR was represented as a lumped resonator, computed from the per-loop capacitance analyzed in prior work~\cite{Sasatani2017} and its measured quality factor.

The QSCR–relay couplings were specified using the link figure of merit $k\sqrt{Q_{\mathrm{TX}}Q_{\mathrm{relay}}}$ for each relay position. For a given relay activation pattern, the loop currents were obtained by solving $\mathbf{V}=\mathbf{Z}\mathbf{I}$ with the QSCR excited by a voltage source. Open relays were modeled by adding a large series resistance to the corresponding relay branch, while retaining the weak direct QSCR–receiver path. The input power (coupled into the QSCR) and the power delivered to each receiver load were then calculated from the solved currents. All $2^5$ relay activation patterns were evaluated, and the delivered power for each receiver was reported as a fraction of the input power.

\subsection*{System implementation}
The room-scale quasistatic cavity resonator was implemented as reported in our prior work~\cite{Sasatani2021}. Briefly, the cavity was constructed from conductive aluminum panels mounted on an aluminum frame, with a central copper pole and lumped capacitors forming the room-scale resonant structure. The drive coil was a $\qty{300}{mm}$-diameter, 10-turn spiral made from AWG 12 copper wire and fixed on an acrylic support. The relay resonator used a $\qty{150}{mm}$ square, six-turn coil made from $\qty{2}{mm}$ diameter copper wire. For the field-bending configuration, two orthogonally oriented relay coils were connected with a coaxial cable; the field-bending simulations used the same six-turn coils as the standard single-relay configuration, whereas the qualitative field-bending demonstration used PCB spiral coils of similar dimensions. The miniature receiver used a $\qty{15}{mm}$-diameter, 10-turn coil made from $\qty{1}{mm}$ diameter copper wire. The measured quality factors of the drive coil, room-scale transmitter, relay resonator, and receiver coil were $Q_{\mathrm{drive}}=396$, $Q_{\mathrm{TX}}=540$, $Q_{\mathrm{relay}}=202$, and $Q_{\mathrm{RX}}=53$, respectively.

All resonant structures were tuned to a common resonant frequency of \qty{1.34}{MHz} using capacitor networks mounted on printed circuit boards. Class~1 NP0/C0G MLCCs were used for coarse tuning because of their low loss and stable capacitance, and variable film capacitors were used for final tuning of the drive coil and relay resonator. The receiver used the series and parallel capacitor network shown in Fig.~\ref{figure:dc-to-dc}, selected so that the impedance seen from the receiver coil matched the target load condition. For the rectified DC measurements, the target AC load was calculated from the standard full-bridge rectifier load transformation, $R_{\mathrm{ac}}=(8/\pi^2)R_{\mathrm{load}}$, with $R_{\mathrm{load}}=\qty{100}{\ohm}$.

AC-to-AC link measurements were performed using a vector network analyzer (Keysight FieldFox N9914A). DC-to-DC measurements were performed by recording the DC input power supplied to the EPC9065 transmitter power stage and the DC power delivered to the load using a bench multimeter.

\subsection*{Use of artificial intelligence}
Generative AI tools (OpenAI GPT, Anthropic Claude, and Google Gemini) were used to refine the language and structure of the manuscript and to assist in developing the analysis code. The authors reviewed and edited all AI-assisted output and take full responsibility for the content of this work.

\section*{Data availability}
The data that support the findings of this study are available at Zenodo
(\url{https://doi.org/10.5281/zenodo.19287418}) and in the project repository
(\url{https://github.com/SasataniLab/hierarchical-resonator-wpt}).

\section*{Code availability}
The code that supports the findings of this study is available at Zenodo
(\url{https://doi.org/10.5281/zenodo.19287418}) and in the project repository
(\url{https://github.com/SasataniLab/hierarchical-resonator-wpt}).

\section*{Competing interests}
The authors declare no competing financial or non-financial interests.

\section*{Acknowledgements}
This work was supported by JST FOREST (Grant Number JPMJFR242P), JST ACT-X (Grant Number JPMJAX190F), and JSPS KAKENHI (Grant Number 23K28068).

\section*{Author contributions}
T.S. designed the research, conceived the experiments, analyzed the results, and wrote the manuscript. A.P.S. and Y.K. reviewed and revised the manuscript. All authors have read and approved the final manuscript.

\bibliography{sample}

\newpage

\clearpage

\clearpage

\clearpage

\clearpage

\clearpage

\end{document}